# Role of pyridine as a biomimetic organo-hydride for homogeneous reduction of CO$_2$ to methanol

Chern-Hooi Lim[1], Aaron M. Holder[1,2], James T. Hynes[2,3], and Charles B. Musgrave[1,2]*

[1]Department of Chemical and Biological Engineering and [2]Department of Chemistry and Biochemistry, University of Colorado, Boulder, Colorado 80309. [3]Chemistry Department, Ecole Normale Supérieure, UMR ENS-CNRS-UPMC 8640, 24 rue Lhomond, 75005 Paris, France

**ABSTRACT:** We use quantum chemical calculations to elucidate a viable homogeneous mechanism for pyridine-catalyzed reduction of CO$_2$ to methanol. In the first component of the catalytic cycle, pyridine (Py) undergoes a H$^+$ transfer (PT) to form pyridinium (PyH$^+$) followed by an e$^-$ transfer (ET) to produce pyridinium radical (PyH$^0$). Examples of systems to effect this ET to populate PyH$^+$'s LUMO (E$^0_{calc}$ ~ -1.3V vs. SCE) to form the solution phase PyH$^0$ via highly reducing electrons include the photo-electrochemical p-GaP system (E$_{CBM}$ ~ -1.5V vs. SCE at pH= 5) and the photochemical [Ru(phen)$_3$]$^{2+}$/ascorbate system. We predict that PyH$^0$ undergoes further PT-ET steps to form the key closed-shell, dearomatized 1,2-dihydropyridine (**PyH$_2$**) species. Our proposed sequential PT-ET-PT-ET mechanism transforming Py into **PyH$_2$** is consistent with the mechanism described in the formation of related dihydropyridines. Because it is driven by its proclivity to regain aromaticity, **PyH$_2$** is a potent recyclable organo-hydride donor that mimics the role of NADPH in the formation of C-H bonds in the photosynthetic CO$_2$ reduction process. In particular, in the second component of the catalytic cycle, we predict that the **PyH$_2$**/Py redox couple is kinetically and thermodynamically competent in catalytically effecting hydride and proton transfers (the latter often mediated by a proton relay chain) to CO$_2$ and its two succeeding intermediates, namely formic acid and formaldehyde, to ultimately form CH$_3$OH. The hydride and proton transfers for the first reduction step, i.e. reduction of CO$_2$, are sequential in nature; by contrast, they are coupled in each of the two subsequent hydride and proton transfers to reduce formic acid and formaldehyde.

## 1. Introduction

Conversion of carbon dioxide (CO$_2$) to fuels enabling a closed-carbon cycle powered by renewable energy has the potential to dramatically impact the energy and environmental fields.[1-10] However, the chemical reduction of CO$_2$ to highly reduced products such as methanol (CH$_3$OH) remains a daunting task. The groups of Fujita,[11-13] Kubiak,[3,14] Meyer,[15-17] Savéant[18-20] and others[21-32] have made significant contributions to this field, particularly in the fundamental understanding of using transition-metal complexes to catalyze CO$_2$'s transformation. Despite these advances, many challenges remain: for example, CO$_2$ reduction has largely been confined to 2e$^-$ products such as CO and formate, and in many cases large overpotentials are required to drive these reactions.[11,14,18,22]

In a striking experimental advance, Bocarsly and coworkers[23,33] identified a promising organic catalyst based on pyridine (Py),[34-36] which is employed photo-electrochemically at a p-type GaP cathode to efficiently convert CO$_2$ to CH$_3$OH at 96% Faradaic efficiency and 300 mV of underpotential.[23] Clearly, thorough understanding of Py-catalyzed CO$_2$ reduction is required not only to elucidate Py's intriguing catalytic role, but also to develop related catalysts that exploit the fundamental phenomena at play in this reduction. In this contribution, we use quantum chemical calculations to discover that the key to Py's catalytic behavior lies in the homogeneous chemistry of the 1,2-dihydropyridine/pyridine redox couple, driven by a dearomatization-aromatization process, in which 1,2-dihydropyridine (**PyH$_2$**) acts as a powerful recyclable organo-hydride that reduces CO$_2$ to CH$_3$OH via three hydride and proton transfer (HTPT) steps (see Scheme 1).

Hydride transfer (HT) reactions --- which are formally equivalent to 2e$^-$/H$^+$ reductions --- have been proven adept in forming C-H bonds, converting CO$_2$ to CH$_3$OH at mild conditions.[24,28,31] For example, we have shown how ammonia borane (H$_3$N-BH$_3$)[37] accomplishes hydride (H$^-$) and proton (H$^+$) transfers to CO$_2$ that ultimately lead to CH$_3$OH.[38,39] The particular relevance of this example is that **PyH$_2$** --- the hydride reagent of special focus in this article and a mimic of biological NADPH --- is similar to ammonia borane in that both involve a protic hydrogen on N which has neighboring hydridic hydrogens, on the ortho-C of 1,2-dihydropyridine and on the B of ammonia borane. However, **PyH$_2$** is unique in the sense that it is a catalytic hydride donor (*vide infra*), similar to NADPH in photosynthesis, rather than a stoichiometric hydride reagent (such as ammonia borane and silanes).

**Scheme 1. Homogeneous reduction of CO$_2$ to methanol by 1,2-dihydropyridine via hydride and proton transfer steps**

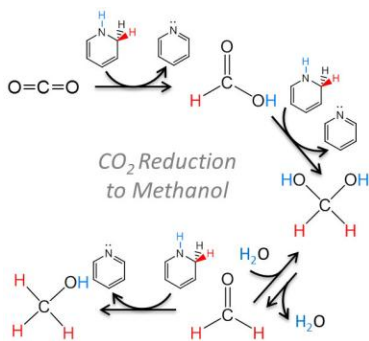

**PyH$_2$** is a 2H$^+$/2e$^-$ transfer product of pyridine (Py).[40-43] It has been observed that the formation of related dihydropyridines (e.g. 1,6-dihydronicotinamide and dihydroacridines) proceeds via sequential proton transfer (PT) and electron transfer (ET) steps.[44-46] In section 3.1, we discuss the initial PT-ET steps where Py is first protonated to form pyridinium (PyH$^+$) and subsequent 1e$^-$ transfer produces pyridinium radical (PyH$^0$). ET to reduce PyH$^+$ to PyH$^0$ involves significant energy input to dearomatize the Py ring.[47] (We show in section 3.8 that dearomatization is crucial in driving catalytic HT reactions that concurrently recover pyridine's aromaticity.) Following the discussion of PyH$^0$ formation, we examine the subsequent PT and ET transfer steps to PyH$^0$ that produce **PyH$_2$** in section 3.2.

The outline of the remainder of this paper is as follows. Using quantum chemical calculations whose methodology is outlined in section 2, we will: 1) demonstrate how Py is transformed into the recyclable organo-hydride 1,2-dihydropyridine, via a sequential PT-ET-PT-ET process (sections 3.1 and 3.2); 2) establish the hydride nucleophilicity of **PyH$_2$** and related dihydropyridines (section 3.3); 3) calculate key transition states to prove that **PyH$_2$** is both kinetically and thermodynamically proficient in homogeneously reducing CO$_2$ to CH$_3$OH through three successive HTPT steps (sections 3.4-3.7); and 4) show that the catalytic hydride transfer reaction by the **PyH$_2$**/Py redox couple is driven by a dearomatization-aromatization process (section 3.8). Concluding remarks are given in section 4.

**2. Computational methods.**

We compute stationary geometries (reactants, transition states and products) for all systems studied using density functional theory based on the M06 density functional[48] and 6-31+G** basis set[49] and a water solvent model described below. The M06 functional was chosen because it has been parameterized with experimental thermodynamic data, and is expected to provide a reliable description of the molecular structures for the reactions of interest here.[48] To further improve the reported energies, we performed single point energy calculations at the M06/6-31+G** geometries using 2$^{nd}$ order Møller-Plesset perturbation theory (MP2)[50] with the extensive aug-ccPVTZ basis sets.[51] We previously found that MP2 accurately reproduces the CCSD(T) reaction and transition state (TS) energies for reactions between pyridine (Py) and CO$_2$,[47] and have further benchmarked this method against CCSD(T) for reactions involving HT to CO$_2$, as summarized in Table S1 of the Supporting Information (SI), section 1.

An adequate treatment of solvent is crucial to correctly describe reactions involving a polar TS, such as those involving electron, proton, or hydride transfers which are of particular interest here. Therefore, we employed the implicit polarized continuum solvation model (CPCM) in all calculations to treat the solute-solvent electrostatic interactions in aqueous solvent.[52,53] In addition to the CPCM-description, in the direct hydride transfer models DHT-1H$_2$O and DHT-2H$_2$O of section 3.3, we explicitly included one and two water molecules to quantum mechanically model the solvent polarization essential for correctly describing the ionic HT TS. In addition to stabilizing the TS, these water molecules also intimately participate in the reaction by acting as a proton relay chain during the proton transfer event.[47,54-67]

We calculate vibrational force constants at the M06/6-31+G** level of theory to: 1) verify that the reactant and product structures have only positive vibrational modes, 2) confirm that each TS has only one imaginary mode and that it connects the desired reactant and product structures via Intrinsic Reaction Coordinate (IRC) calculations, and 3) compute entropies, zero-point energies (ZPE) and thermal corrections included in the reported free energies at 298K.

For the activation and reaction enthalpies, entropies and free energies for each of the various reactions examined within, we define the reference state as the separated reactants, as is appropriate for bimolecular reactions.[68] It is important to recognize that commonly employed entropy evaluations within the rigid rotor, harmonic oscillator and ideal gas approximations normally overestimate the entropic cost for reactions occurring in solution phase, because ideal gas partition functions do not explicitly take into account hindered translation, rotation and vibration of the solute surrounded by solvent molecules.[25,69-74] For example, Huang and coworkers observed that the calculated standard activation entropy values (-T$\Delta$S$^‡_{calc}$) consistently overestimate the experimental -T$\Delta$S$^‡_{exp}$ values by ~4-5 kcal/mol at 298K.[71,72] Liang and coworkers also observed that -T$\Delta$S$^‡_{exp}$ values are 50-60% of the computed -T$\Delta$S$^‡_{calc}$, and in some cases activation entropic costs -T$\Delta$S$^‡_{exp}$ are overestimated by ~11 kcal/mol.[70] In SI, section 2, we show that -T$\Delta$S$^‡_{calc}$ overestimates -T$\Delta$S$^‡_{exp}$ by ~12 kcal/mol for the analogous HT reaction from the **PyH$_2$**-related dihydropyridine 1-benzyl-1,4-dihydronicotinamide (in eq. 1). Clearly, ideal gas-based calculated -T$\Delta$S$^‡_{calc}$ values can have significant errors.

While various empirical correction factors for $-T\Delta S^{\ddagger}_{calc}$ values have been proposed,[25,69,74,75] all of which significantly lower $-T\Delta S^{\ddagger}_{calc}$, our approach to better estimate $-T\Delta S^{\ddagger}$ is to employ the experimentally obtained $-T\Delta S^{\ddagger}_{exp}$ value for an analogous HT reaction; as we discuss later, the transition states for all three steps in reduction of $CO_2$ to $CH_3OH$ are of HT character. This $-T\Delta S^{\ddagger}_{exp}$ value is then added to our calculated $\Delta H^{\ddagger}_{HT}$ in order to obtain more accurate estimates to the activation free energy $\Delta G^{\ddagger}_{HT}$. In particular, the homogeneous HT from the **PyH$_2$**-related dihydropyridine 1-benzyl-1,4-dihydronicotinamide to $\Delta^1$-pyrroline-2-carboxylic acid (zwitterionic form) in aqueous methanol (eq. 1)[76] is analogous to each of the three HTs from **PyH$_2$** of interest here: to $CO_2$, formic acid (HCOOH) and formaldehyde ($OCH_2$). We thus add the $-T\Delta S^{\ddagger}_{exp}$ of 2.3 kcal/mol (298 K) determined experimentally for eq. 1[76] to the calculated $\Delta H^{\ddagger}_{HT}$ values in Table 1 to obtain our estimates for $\Delta G^{\ddagger}_{HT}$. This procedure is further discussed in section 3.5. As comparison, we also employed Morokuma and coworkers' approach[77] to omit the translational contribution from computed gas phase entropies. We obtained $-T\Delta S^{\ddagger}_{calc}$ = 3.0, 2.2, and 2.7 kcal/mol for the reduction of $CO_2$, formic acid and formaldehyde, respectively, via the DHT-1H$_2$O model; these values are similar to the experimental $-T\Delta S^{\ddagger}_{exp}$ of 2.3 kcal/mol for eq.1 that we employed. See SI, section 2 for details.

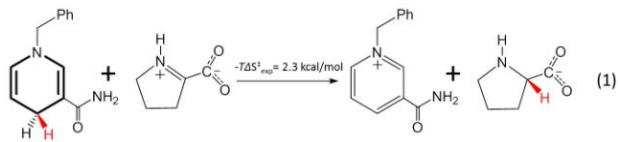

Finally, reaction free energies ($\Delta G^0_{rxn}$) were reported by adding $\Delta H^0_{rxn}$ to $-T\Delta S^0_{rxn}$ in Table 1. Because the number of species remains constant on going from reactants to products in the HTPT reactions described here, the overestimation issue for the calculated $-T\Delta S^0_{rxn}$ is less severe. All reported energies were referenced to separated reactants (as noted above) and calculations were performed using the GAUSSIAN 09[78] and GAMESS[79] computational software packages. Often, reported bimolecular reaction activation and thermodynamic quantities in the literature are referenced to reactants within a reactant complex rather than to the separated reactants. Thermodynamic quantities with the former reference are given for comparison in SI, section 3.

### 3. Results and Discussion

**3.1 Formation of PyH$^0$ from Py via 1H$^+$/1e$^-$ transfers.** We begin with the key issue of the generation of PyH$^0$ from Py via sequential PT-ET steps. In Scheme 2, route **I**, Py first undergoes protonation to form pyridinium (PyH$^+$; pK$_a$= 5.3) in a pH= 5 solution. Subsequent 1e$^-$ reduction (route **II**) produces PyH$^0$. Experimentally, photo-excited electrons of the p-GaP semiconductor are sufficiently reducing to populate PyH$^+$'s LUMO ($E^0_{calc}$~ -1.3 V vs. SCE)[47,80,81] via 1e$^-$ transfer to form solution phase PyH$^0$.[82] For example, at a pH of 5 the conduction band minimum of p-GaP ($E_{CBM}$)[83,84] lies at approximately -1.5 V vs. SCE,[85,86] a more negative potential than PyH$^+$'s LUMO. Furthermore, the p-GaP electrode is electrochemically biased by -0.2 to -0.7 V,[23] which further increases the reducing ability of the transferring electron. PyH$^0$ can also be produced electrochemically at inert electrodes. For instance, a glassy carbon electrode[87,88] has been used to electrochemically produce similar neutral radicals from the Py-related species nicotinamide and acridines.[44-46] In another case, photochemical production of PyH$^0$ driven by visible light was recently demonstrated by MacDonnell and coworkers using a surface-free photochemical process in which Ru(II) trisphenanthroline (chromophore) and ascorbate (reductant) act in concert to reduce PyH$^+$ to PyH$^0$ via 1e$^-$ transfer. The produced PyH$^0$ radical is actively involved in the observed homogeneous reduction of $CO_2$ to $CH_3OH$,[89,90] an observation in contrast with recent studies focused on the specific case using a Pt cathode[81,87,91-95] that rule out participation of homogenous PyH$^0$ in Py-catalyzed $CO_2$ reduction. We stress that we consider a Pt electrode to be a special case. On a Pt electrode, 1e$^-$ reduction of PyH$^+$ is favored to form adsorbed H-atoms (Pt-H*)[93-96] such that using it as a cathode introduces additional routes (e.g. $H_2$ formation) which likely outcompete any processes catalyzed by Py. Therefore, surface pathways[92,94] for $CO_2$ reduction on Pt may predominate such that the homogeneous mechanism discussed in the text that requires the production of PyH$^0$ becomes a minor pathway.

**Scheme 2. Formation of pyridinium radical (PyH$^0$)**

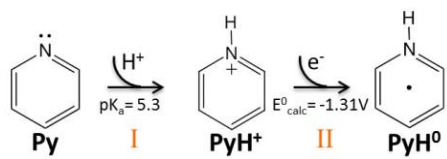

The conversion of the produced solution phase PyH$^0$ to the desired intermediate **PyH$_2$** will be taken up in section 3.2. Here we pause to discuss some competing routes. The first of these arises because PyH$^0$ is a dearomatized species driven to donate an electron in order to recover its aromaticity.[47,97] For example, Bocarsly and coworkers[33,98] proposed that PyH$^0$ reacts with $CO_2$ to form a pyridine-carbamate (PyCOOH$^0$) intermediate (Scheme 3, route **III**) prior to $CH_3OH$ formation.[33] Formation of PyCOOH$^0$ by this route is supported by our recent computational study,[47] and by spectroscopic measurements.[99] In particular, using a hybrid explicit/implicit solvent model, we calculated low enthalpic barriers with respect to the complexed reactants of 13.6-18.5 kcal/mol (depending on the number of solvating waters) for PyCOOH$^0$ formation via a proton relay mechanism; the importance of proton relays have been extensively described in assorted chemical reactions.[54-65] Charge analysis on $CO_2$ and PyH$^0$ along the reaction coordinate reveals that PyH$^0$'s propensity to recover its aromaticity drives the sequence of ET to $CO_2$ followed by PT (mediated by a proton relay) to ultimately form PyCOOH$^0$.[47,100] While this particular reaction is not of direct interest in the present work,[101] we will see that the themes of aromaticity recovery and proton relay mechanisms also prove to be important for our three HTPT step reduction of $CO_2$ to $CH_3OH$.

Another oxidation channel for PyH⁰ is via radical self-quenching, shown in route **IV**. PyH⁰ undergoes self-quenching[102] to form either H₂ + 2Py or a 4,4' coupled dimer;[96,103] the recovery of Py catalyst from the 4,4' coupled dimer is demonstrated in SI, section 4. Interestingly, the self-quenching of PyH⁰ can also lead to a productive outcome: disproportionation[104] of two PyH⁰ radicals leads to Py and the desired **PyH₂** species.[105] However, we consider that the main route to **PyH₂** is not disproportionation, but a successive PT and ET to PyH⁰,[106,107] which we now describe.

**Scheme 3. 1e⁻ reduction of CO₂ by PyH⁰ to form PyCOOH⁰ and self-radical quenching reactions of PyH⁰**

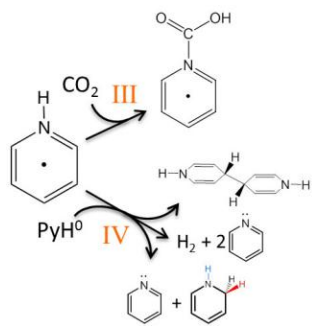

**3.2 Formation of 1,2-dihydropyridine (PyH₂) from PyH⁰ via successive 1H⁺/1e⁻ transfers.** We now discuss production of **PyH₂** from PyH⁰ via routes **V** and **VI** of Scheme 4 in which PyH⁰ undergoes further 1H⁺ and 1e⁻ transfers to form closed-shell solution phase **PyH₂**. We propose that these routes are competitive with, if not predominant over, Scheme 3's quenching routes **III** and **IV**. In particular, given that quenching routes are second-order in [PyH⁰] and that routes **III** and **V** are first-order in [PyH⁰], it is likely that quenching would prevent the concentration of PyH⁰ from reaching a level at which the second-order process dominates. Furthermore, a significant fraction of any self-quenching of PyH⁰ that does occur could lead to the desired **PyH₂** species, as observed experimentally for quenching of the related 3,6-diaminoacridinium radical to form the corresponding dihydropyridine species (3,6-diaminoacridan).[104,105]

**Scheme 4. Formation of 1,2-dihydropyridine (PyH₂)**

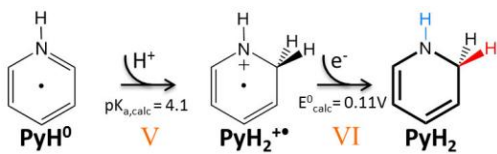

The importance of protonation of PyH⁰ by our proposed route **V** depends critically on the rate of PT to PyH⁰, which we now address in some detail. The $pK_a$ of PyH₂⁺· is only 4.1 (at the C₂ carbon),[108-110] indicating that at a pH of 5, only a fraction of PyH⁰ is protonated in solution. However, PyH⁰ is produced by reduction of PyH⁺ at the cathode near the double layer region where the lower pH facilitates its protonation to form PyH₂⁺·.[111] The key here is that near the double layer region, the electric field created by the applied negative bias at the cathode concentrates cationic PyH⁺ and H₃O⁺ species according to a Poisson-Boltzmann distribution,[112-114] significantly lowering the pH near the cathode surface. For example, in SI section 5, we use a linearized Poisson-Boltzmann model to demonstrate that the concentration of cation acids, e.g. H₃O⁺ and PyH⁺, increases by several orders of magnitude as they approach the negatively biased cathode. Thus, protonation of PyH⁰ by PyH⁺ or H₃O⁺ near the cathode double layer to form the radical cation PyH₂⁺· becomes a highly probable event. It has a much higher probability than does radical self-quenching via route **IV** because [cation acids]>>[PyH⁰], and as described above, protonation is first-order in [PyH⁰], while quenching is second-order in [PyH⁰]. The absence of a cathode double layer effect is consistent with the observations by MacDonnell and coworkers in their homogeneous Ru(II)/ascorbate photochemical system in which cationic and PyH⁰ species are dispersed uniformly in the solution in the absence of a cathode. We argue that this effect concentrates cations at the productive surface in Bocarsly's system. The impact of its absence is also consistent with MacDonnell and coworkers' observation that high PyH⁺/Ru(II) ratios of ∼100 were required to produce CH₃OH, which we propose is required to drive protonation of PyH⁰ in the absence of a cathode.[89]

Finally, **PyH₂** is produced by reduction of PyH₂⁺· in proposed route **VI** in Scheme 4. Our calculated positive reduction potential for PyH₂⁺· of $E^0_{calc}$= 0.11 V vs. SCE indicates that PyH₂⁺· reduction is facile and consequently that 1e⁻ transfer (from PyH⁰ or via a photoexcited electron) to PyH₂⁺· to form **PyH₂** is realized on p-GaP and in MacDonnell and coworkers' homogeneous Ru(II)/ascorbate photochemical system.[115] Our suggested sequential PT-ET-PT-ET sequence (Schemes 2 and 4, route **I-II-V-VI**) to form **PyH₂** from Py is strongly supported by the fact that an analogous process has been observed for the conversion of the Py-related species nicotinamide,[44,116] acridine,[45,117] and 3,6-diaminoacridine (proflavine)[46] to their related dihydropyridine species. We note that we propose the formation of 1,2-dihydropyridine as the kinetic product[40] because protonation of the PyH⁰'s C2 carbon is more facile than protonation at the 4 position,[108] analogous to protonation of nicotinamide where the related 1,2-dihydropyridine is formed.[44] However, 1,4-dihydropyridine can also be produced, although at a slower rate.[40] In SI, section 6, we show both **PyH₂** species to be capable of direct HT, with 1,2-dihydropyridine being the slightly more reactive species. We also note that acid-catalyzed hydration of both 1,2-dihydropyridine and 1,4-dihydropyridine may generate undesirable side products.[118,119]

We have thus far described likely steps that transform Py into **PyH₂**, a species which we now show to be competent in performing catalytic direct HT to carbonyls.

**3.3 Establishing the hydride nucleophilicity of PyH₂ and related dihydropyridines.** First, it is noteworthy that **PyH₂** chemically resembles the NADPH dihydropyridine species found in nature (see Scheme 5a and caption) that acts in the NADPH/NADP⁺ redox cycle of photosynthesis to produce sugars from CO₂ by hydride transfers.[120,121] In-deed, since the discovery of NADPH in the 1930's, related dihydropyridine compounds have been studied, especially in connection with their HT to various substrates containing C=C, C=N and C=O groups.[40-43] HT to carbonyls is obviously of particular interest here: the reactant $CO_2$ and its reduced intermediates formic acid (HCOOH) and formaldehyde ($OCH_2$) leading to $CH_3OH$ formation all contain C=O groups susceptible to HT.

**Scheme 5. Reductions via direct hydride transfers from related dihydropyridines species**

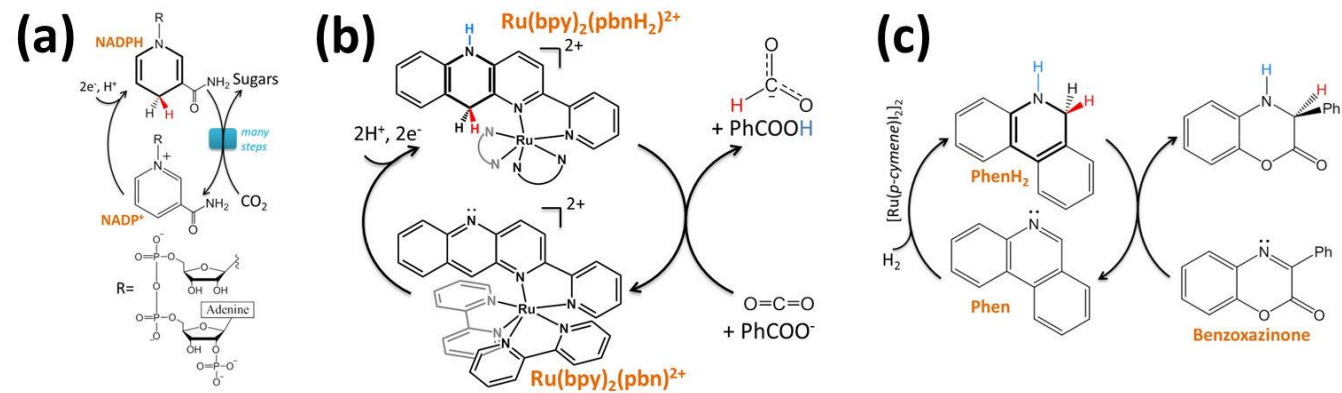

(**a**) NADPH/NADP[+] redox cycle of photosynthesis to produce sugars from $CO_2$ by hydride transfers. NADPH creates a C-H bond by HT to a carbonyl group --- not in $CO_2$ --- in a key reduction in the multi-step photosynthetic process. (**b**) Catalytic reduction of $CO_2$ to formate via HT involving Tanaka's Ru-based dihydropyridine species (**Ru(bpy)$_2$(pbnH$_2$)$^{2+}$**); bpy= 2,2'-bipyridine, pbn= 2-(pyridin-2-yl)benzo[b][1,5]naphthyridine).[29,122] (**c**) Catalytic hydrogenation (via hydride and proton transfer) of benzoxazinone by Zhou's dihydrophenanthridine species (**PhenH$_2$**).[123]

Here we mention two examples of related recyclable dihydropyridines performing HT to the C=O and C=N groups. Tanaka and coworkers demonstrated[122] (Scheme 5b) that the electrochemical reduction of Ru(bpy)$_2$(pbn)$^{2+}$ forms the NADPH-like **Ru(bpy)$_2$(pbnH$_2$)$^{2+}$**, where the pbn ligand has undergone 2H[+]/2e[-] transfer to form a dihydropyridine-like hydride donor.[124] Association of **Ru(bpy)$_2$(pbnH$_2$)$^{2+}$** with a benzoate base (PhCOO[-]) then activates its hydride donation to $CO_2$ to form HCOO[-] and PhCOOH and to concomitantly regenerate Ru(bpy)$_2$(pbn)$^{2+}$.[29] An H/D kinetic isotope effect of 4.5 was determined, further supporting the direct hydride transfer mechanism to $CO_2$ to form HCOO[-].[29] Similarly, Zhou et al's dihydrophenanthridine (**PhenH$_2$**), a **PyH$_2$** analog, catalytically transfers both its hydride and proton to benzoxazinone and regenerates the phenanthridine catalyst (Scheme 5c), further demonstrating the competence of dihydropyridine species as recyclable hydride donors.[123]

We have thus far argued that the HT reactivity of related dihydropyridine hydrides **NADPH**, **Ru(bpy)$_2$(pbnH$_2$)$^{2+}$** and **PhenH$_2$** --- especially the extraordinary ability of **Ru(bpy)$_2$(pbnH$_2$)$^{2+}$** to effect $CO_2$ reduction --- strongly implicates **PyH$_2$** as a robust hydride donor in Bocarsly and coworkers' Py-catalyzed $CO_2$ reduction. The next natural step is to quantify **PyH$_2$**'s ability as a hydride donor, i.e. its hydride nucleophilicity. Figure 1 shows the quantification of the hydride nucleophilicity of hydride donors using Mayr and coworkers' Nucleophilicity (N) values,[125,126] where large N values indicate strong hydride donor ability. Note that the N scale is a kinetic parameter quantifying the HT rate, whereas the often-employed hydricity is a thermodynamic parameter.[127-129] In order to place the N values of **PyH$_2$** and Zhou's **PhenH$_2$** in perspective relative to established values for dihydropyridines and NaBH$_4$, we cal-culate activation free energies for HT ($\Delta G^{\ddagger}_{HT}$) from these donors to $CO_2$ to reduce it to formate (HCOO[-]) via the Direct-Hydride-Transfer (DHT) model illustrated in Figure 2a.

In Figure 1, we use the experimental N and our calculated $\Delta G^{\ddagger}_{HT}$ values (in kcal/mol) of 1,4-cyclohexadiene (0.09, 53.0), 10-methyl-9,10-dihydroacridine (5.54, 40.5), Hantzsch's ester (9.00, 29.9), and NaBH$_4$ (14.74, 13.8) to obtain a nearly linear relationship between $\Delta G^{\ddagger}_{HT}$ and N: $\Delta G^{\ddagger}_{HT}$ = -2.70*N + 54.1.[130] We then use this linear relation together with our calculated $\Delta G^{\ddagger}_{HT}$ barriers to estimate that the N values of **PhenH$_2$** and **PyH$_2$** are 8.1 and 11.4, respectively. Although **PyH$_2$** is a less capable hydride donor than the well-known strong donor NaBH$_4$, it is the most reactive dihydropyridine, reducing $CO_2$ to HCOO[-] at $\Delta G^{\ddagger}_{HT}$ = 23.2 kcal/mol by the DHT model. The hydricity of **PyH$_2$** was also calculated according to Muckerman et al.'s approach;[129] we obtained a value of 41.5 kcal/mol (< 43 kcal/mol of HCOO[-]), which supports that HT from **PyH$_2$** to $CO_2$ is thermodynamically favorable.[131]

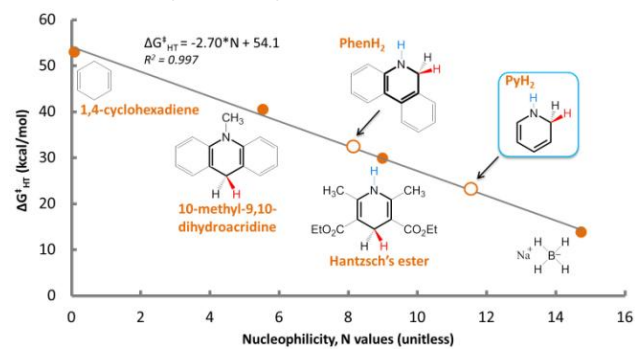

**Figure 1.** The activation free energy of hydride transfer to $CO_2$ varies linearly with hydride nucleophilicity. $\Delta G^\ddagger_{HT}$ (kcal/mol) is our calculated activation free energy for direct HT (DHT) to $CO_2$ to form $HCOO^-$. The activation free energy $\Delta G^\ddagger_{HT}$ is obtained by adding our calculated $\Delta H^\ddagger_{HT}$ to the experimental $-T\Delta S^\ddagger_{exp}$ = 2.3 kcal/mol for the analogous HT reaction eq. 1, with all quantities referenced to the separated reactants (see section 2). Nucleophilicity (N) values quantify the strength of hydride donors.[125,126] The equation log k(20°C) = s(N+E) was used to obtain N and s (the slope factor) values in order to generalize various classes of hydride donors, including dihydropyridines and borohydrides. HT rate constants k are measured at 20°C for HT to acceptors with known E (Electrophilicity) values. Our calculated $\Delta G^\ddagger_{HT}$ values are used to estimate k, and thus N values of **PyH$_2$** and Zhou's **PhenH$_2$** relative to established N values for dihydropyridines and NaBH$_4$. These $\Delta G^\ddagger_{HT}$ values are obtained with $CO_2$ acting as the hydride acceptor; $CO_2$'s E value is unknown but this is immaterial to the estimation of **PyH$_2$** and **PhenH$_2$**'s N values.[132] The comparatively low $\Delta G^\ddagger_{HT}$ and high hydride nucleophilicity values for **PyH$_2$** are apparent in this Figure.

With these important preliminaries concerning **PyH$_2$**'s generation and HT ability concluded, we now turn to the three HTPT steps in the reduction of $CO_2$ to methanol.

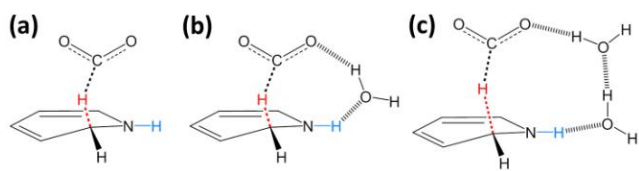

**Figure 2.** Hydride transfer to $CO_2$ can occur through various direct hydride transfer configurations. Here, we model three possible HT configurations, without (a) and with (b and c) the active participation of H$_2$O, which we demonstrate are kinetically and thermodynamically favorable towards reducing $CO_2$: (a) Direct-Hydride-Transfer (DHT) model, (b) DHT-1H$_2$O model where one H$_2$O acts as a proton relay and (c) DHT-2H$_2$O model where two H$_2$O's act as a proton relay. Details of these relays are discussed subsequently.

**3.4 First HTPT step: PyH$_2$ + CO$_2$ → Py + HCOOH.** We now elaborate the first HTPT step in $CO_2$'s conversion to $CH_3OH$: HT to $CO_2$ by **PyH$_2$** to form formic acid (HCOOH). This step is illustrated in Scheme 6, route **VII**, although as we will see, there are two sequential steps involved, namely first formate ion HCOO$^-$ production followed by formic acid generation.[133] The $\Delta G^\ddagger_{HT}$ for this step occurring without the electrostatic effects and active participation of the proton relay (predicted using the DHT model in Figure 2a) is 23.2 kcal/mol. This shows that even without the effects described by explicit water, HT is kinetically viable.

**Scheme 6. Reduction of CO$_2$ to formic acid by PyH$_2$.**

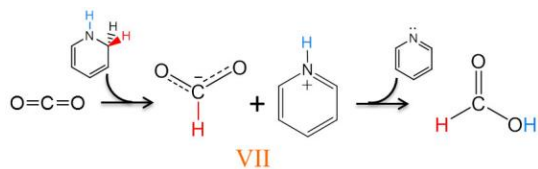

In an attempt to improve the description beyond the DHT model, we have considered two likely elaborations in aqueous solution. We added one and two solvating water molecules (DHT-1H$_2$O and DHT-2H$_2$O, Figures 2b and c) to polarize the reactive complex beyond the polarization afforded by implicit solvent, and thus stabilize the ionic TS relative to the neutral reactants. As will be seen, in the formic acid and formaldehyde reductions, the solvating water molecule(s) play an additional, more active role; they act as a proton relay, for which this mixed explicit/implicit solvation approach[59,60,134] is especially important for an accurate description.[47,54-57] For the DHT-1H$_2$O and DHT-2H$_2$O models, we obtain the barriers of $\Delta G^\ddagger_{HT}$ = 17.1 and 14.3 kcal/mol for the $CO_2$ reduction to HCOO$^-$, ~6 and 9 kcal/mol lower than for the DHT model, reflecting the importance of quantum mechanically described water polarization (see Table 1).

**Table 1. Activation and reaction free energies and enthalpies for HTPT steps from PyH$_2$ to CO$_2$, HCOOH and OCH$_2$ via various HT models in Figure 2.**

| Model[a] | CO$_2$[b] | | HCOOH[c] | | OCH$_2$[d] | |
|---|---|---|---|---|---|---|
| | $\Delta G^\ddagger_{HT}$ ($\Delta H^\ddagger_{HT}$) | $\Delta G^0_{rxn}$ ($\Delta H^0_{rxn}$) | $\Delta G^\ddagger_{HT}$ ($\Delta H^\ddagger_{HT}$) | $\Delta G^0_{rxn}$ ($\Delta H^0_{rxn}$) | $\Delta G^\ddagger_{HT}$ ($\Delta H^\ddagger_{HT}$) | $\Delta G^0_{rxn}$ ($\Delta H^0_{rxn}$) |
| DHT | 23.2 (20.9) | -9.2 (-5.5) | 25.6 (23.3) | -12.8 (-12.8) | 14.5 (12.2) | -31.3 (-31.4) |
| DHT-1H$_2$O | 17.1 (14.8) | -8.3 (-10.8) | 23.4 (21.1) | -10.6 (-10.8) | 8.9 (6.6) | -31.9 (-31.8) |
| DHT-2H$_2$O | 14.3 (12.0) | -5.6 (-9.8) | 18.7 (16.4) | -11.9 (-12.2) | 6.0 (3.7) | -30.8 (-31.9) |

[a]All free energies and enthalpies, referenced to separated reactants, are reported in kcal/mol at 298K and 1 atm. 2e$^-$/2H$^+$ transfer products: [b]formic acid, [c]methanediol and [d]methanol. The $CO_2$ pathway involves a sequential HT (to produce formate) followed by PT (to produce formic acid); the activation barriers displayed refer to the HT portion of the reaction. The formic acid and formaldehyde reduction pathways both involve a coupled HTPT process, where **PyH$_2$** transfers both its hydridic and protic hydrogens to HCOOH and OCH$_2$, respectively: each case involves a single TS of HT character, with the PT following at a slightly later time, without a separate TS. The formaldehyde reduction step is preceded by the dehydration of methanediol to formaldehyde ($K_{eq}$ = ~5x10$^{-4}$); see Figure 3 and section 3.6. Calculated imaginary frequencies corresponding to the transition state structures are reported in the SI, section 8.

Analysis of the reaction path using an IRC calculation shows that the TS is of HT character, such that the use of the experimental HT activation entropy factor discussed at the end of section 2 is appropriate.[135] The IRC analysis also shows that the product complex consists of the formate anion HCOO[-] and PyH[+]; the reaction is pure HT without any PT, even with a proton relay chain of one or more explicit water molecules included. Because HCOOH's $pK_a$ of 3.8 is relatively low, the carbonyl of HCOO[-] is not basic enough to abstract a proton from its neighboring H-bonded water to initiate a proton relay that would effectively transfer the proton from PyH[+] to HCOO[-]. On the other hand, in sections 3.5 and 3.6, we will show that the HT intermediary products of formic acid (hydroxymethanolate, (HCOOH)H[-]) and formaldehyde (methoxide, OCH$_3$[-]) are highly basic and do initiate a proton relay; PyH[+]'s proton is effectively transferred to these species through the proton relay to form methanediol and methanol, respectively.

Thus, with all three models, the formate product remains unprotonated. However, for the next HTPT step to proceed, HCOO[-] must first be protonated to form formic acid (HCOOH). Yet HCOOH's low $pK_a$ of 3.8 indicates that at equilibrium and a pH = 5 only a small fraction (~1/16 at 298 K) of HCOO[-] is protonated to produce HCOOH. Nevertheless, as described in section 3.2, the high [H$_3$O[+]] and [PyH[+]] near the cathode surface[33,112] (as well as the high [PyH[+]] used in the homogeneous Ru(II)/ascorbate photochemical system[89]) increases the concentration of HCOOH in equilibrium with HCOO[-]. Thus, the first HTPT step to reduce CO$_2$ is sequential, with HT (to produce a relatively stable HCOO[-] intermediate corresponding to a minimum on the HT potential energy surface) followed by a subsequent PT (to produce HCOOH), which we write collectively as **PyH$_2$** + CO$_2$ → Py + HCOOH. We could also term this step-wise HTPT as *uncoupled* HTPT.

Py and HCOOH formation by **PyH$_2$** + CO$_2$ → Py + HCOOH with all three DHT models have negative reaction free energies $\Delta G^0_{rxn}$ of ~ -9 to -6 kcal/mol as shown in Table 1. This demonstrates that **PyH$_2$** is both kinetically and thermodynamically competent in catalytically reducing CO$_2$, at least for the first HTPT step. We will show that this catalytic ability also holds for the remaining two HTPT steps to attain methanol. The schematic free energy surface for this first HTPT step to transform CO$_2$ into HCOOH is shown in Figure 3, which also illustrates the energetics of the two subsequent HTPT steps described in sections 3.5 and 3.6.

We close the discussion of this first CO$_2$ reduction step with two remarks. First, although we have considered only three models (Figures 2a-c) for HT from **PyH$_2$** to CO$_2$, other configurations --- such as DHT-K[+] and DHT-PyH[+] where a potassium cation (present as an electrolyte) and the pyridinium cation act as a Lewis acid and a Brønsted acid, respectively, to activate and stabilize HT[136] to CO$_2$ --- can also lead to the desired HCOOH and Py products. Furthermore, because the reaction is carried out in aqueous solvent, we propose that DHT-1H$_2$O, DHT-2H$_2$O and other likely DHT models with somewhat longer water proton relay chains contribute significantly to the ensemble-weighted average $\Delta G^{\ddag}_{HT}$. Secondly, all reported $\Delta G^{\ddag}_{HT}$ values in Table 1 (including $\Delta G^{\ddag}_{HT}$ for the first HTPT step to form HCOOH and Py) are derived by adding our calculated $\Delta H^{\ddag}_{HT}$ to the experimentally obtained $-T\Delta S^{\ddag}_{exp}$ = 2.3 kcal/mol for an analogous HT reaction eq. 1 (again, all quantities are referenced to separated reactants). This is a more reliable estimate for solution phase HT from **PyH$_2$** than a calculated $-T\Delta S^{\ddag}_{calc}$ based on ideal gas assumptions, which can severely overestimate the entropic contribution to $\Delta G^{\ddag}$;[25,69-74] see section 2 for a more detailed discussion.

**3.5 Second HTPT step: PyH$_2$ + HCOOH → Py + CH$_2$(OH)$_2$.** We now turn to the second HTPT step: the reduction of formic acid to methanediol (CH$_2$(OH)$_2$), as illustrated in Scheme 7, route **VIII**. HCOOH's reduction is actually more challenging than that of CO$_2$, a feature implied by the fact that most CO$_2$ reduction catalysts produce HCOO[-]/HCOOH, but fail to convert HCOOH to more reduced products.[11,14,18] A further indication is provided by the observations of MacDonnell and coworkers, who found a significant build-up of HCOO[-], reflecting the challenge of HCOOH reduction.[89] The key characteristic of HCOOH that makes it difficult to reduce is its highly negative electron affinity (EA); we calculated the gas phase adiabatic EA of HCOOH to be -1.22 eV, which is significantly more negative than the -0.60 eV EA of CO$_2$ (see SI, section 1c) and indicates that, as noted above, formic acid is even more challenging to reduce than CO$_2$.[137,138] We now examine **PyH$_2$**'s ability to reduce HCOOH.

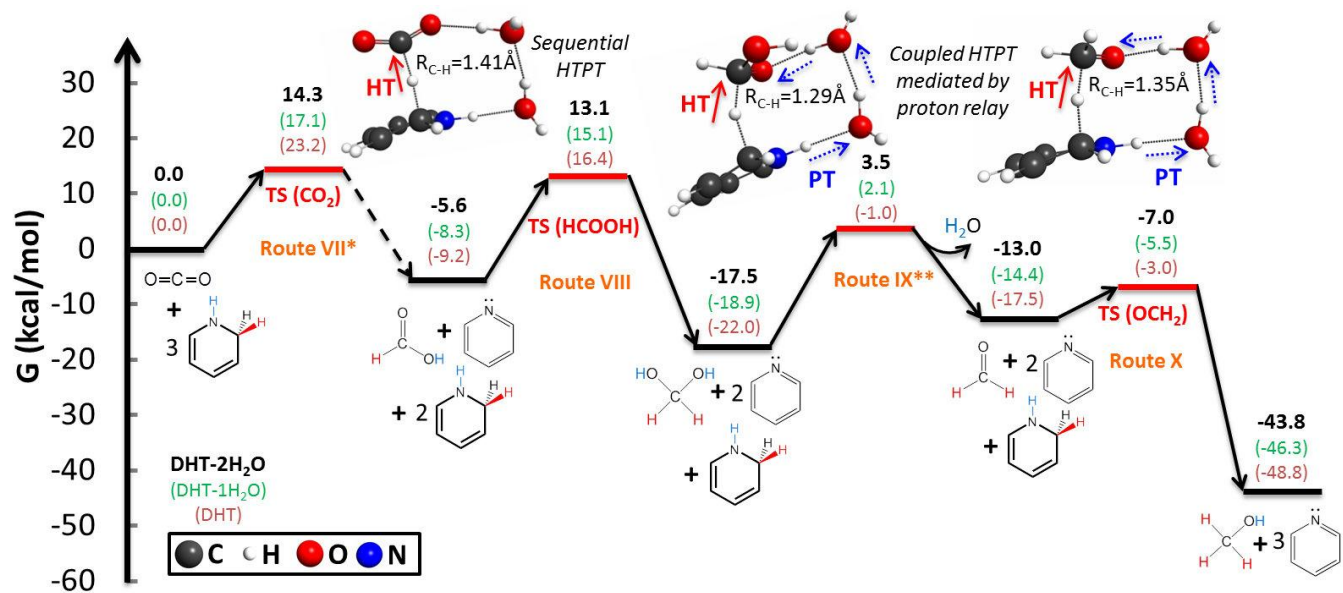

**Figure 3.** Conversion of $CO_2$ to $CH_3OH$ and $H_2O$ by **PyH$_2$** proceeds through three hydride and proton transfer steps. The reported free energies correspond to stationary points along the reaction potential energy surface using the DHT-2H$_2$O (Black), DHT-1H$_2$O (Green) and DHT (Orange) models, catalyzed by HTPT reactions of the **PyH$_2$**/Py redox couple. The 1$^{st}$ HTPT step (Scheme 6, route **VII**) is sequential where HT from **PyH$_2$** to $CO_2$ forms stable formate ($HCOO^-$), with a single TS of HT character, and subsequent PT follows to produce formic acid (HCOOH); (*the dashed line indicates that the product of HT to $CO_2$ is formate where a separate protonation step is required to form formic acid.) In the 2$^{nd}$ HTPT step (Scheme 7, route **VIII**), coupled HTPT occurs with a single TS: HT from **PyH$_2$** to HCOOH, which dominates the barrier and is followed by PT without an additional TS (from oxidized **PyH$_2$**, essentially a PyH$^+$) is mediated by a proton relay involving water molecules, ultimately producing methanediol ($CH_2(OH)_2$). Prior to the next reduction step, $CH_2(OH)_2$ is dehydrated to form the reactive formaldehyde ($OCH_2$) species at $K_{eq}$ ~$5\times10^{-4}$ (Scheme 8, route **IX**); thus this constitutes an additional free energy activation cost of ~4.5 kcal/mol for $OCH_2$ reduction. (**The rate constant for the dehydration of $CH_2(OH)_2$ to $OCH_2$ at ambient conditions[139,140] is ~$5\times10^{-3}$ s$^{-1}$ or equivalently the estimated $\Delta G^\ddagger_{dehyd}$ is ~21 kcal/mol. Consequently, the effective rate constant for transformation of $CH_2(OH)_2$ to $CH_3OH$ is that of $CH_2(OH)_2$ dehydration.) In the 3$^{rd}$ and final HTPT step (Scheme 8, route **X**), which is similar to HCOOH reduction, coupled HTPT occurs where HT from **PyH$_2$** to $OCH_2$, involves a single TS of HT character, and is followed by a proton relay-mediated PT without an additional TS to ultimately form methanol ($CH_3OH$). During each reaction step, the Py catalyst is recovered, thus demonstrating the **PyH$_2$** as a recyclable organo-hydride. TS structures for the HTPT steps from **PyH$_2$** to $CO_2$, HCOOH and $OCH_2$ are shown for the DHT-2H$_2$O model. (Coordinates for the TS structures for all three DHT models are reported in SI, section 9.) All TS structures are HT in character. Animations of the HTPT steps (via DHT-1H$_2$O) for the reduction of $CO_2$, HCOOH and $OCH_2$ are available in the HTML version of the paper.

Table 1 summarizes both $\Delta G^\ddagger_{HT}$ and $\Delta G^0_{rxn}$ for the second HTPT step: **PyH$_2$** + HCOOH → Py + $CH_2(OH)_2$ via the three HT models shown in Figures 2a-c; note that the $CO_2$ 4e$^-$ reduction product methanediol is formed along with the recovery of the Py catalyst. The $\Delta G^\ddagger_{HT}$ of 23.4 kcal/mol for the DHT-1H$_2$O case is ~2 kcal/mol lower than the DHT barrier (25.6 kcal/mol), while the DHT-2H$_2$O model reaction results in a further lowering of $\Delta G^\ddagger_{HT}$ to 18.7 kcal/mol (see Figure 3 for the computed TSs for the DHT-2H$_2$O model). As we will soon see, this reduction only involves a single TS and is thus a *coupled* HTPT process. The character of the TS is primarily that of HT, with PT occurring subsequently without its own TS (as implied in Figure 4, to be discussed). This supports our use of the HT activation entropy factor of section 2. In fact, because the PT occurs along the exit channel ~10-15 kcal/mol below the TS, even an unusually large -T$\Delta S^\ddagger$ for PT would not limit the rate of HTPT.

**Scheme 7. Reduction of formic acid to methanediol by PyH$_2$.**

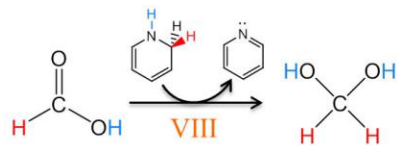

The DHT model results with one and two explicit waters show that HCOOH reduction to generate $CH_2(OH)_2$ is aided by a proton relay chain involving explicit water. Such chains of course stabilize the ionic TS, but they also facilitate PT by reducing strain in the TS and in addition, the PT from the $H_2O$ H-bonded to HCOOH (see Figure 4) stabilizes the partially reduced product as negative charge accumulates on HCOOH. Consequently, the coupled PT helps to overcome the reduction challenges associated with HCOOH's low EA.

This PT and subsequent PTs in the relay chain occur after the HT barrier (see Figure 4a) and of course before the stable products are formed (see Figure 4 for the DHT-1H$_2$O case). Only a very modest activation entropy effect is anticipated here because in the coupled HTPT process, the PT step(s) is (are) considerably delayed relative to the HT such that any entropic penalties due to PT contribute to the free energies of structures well past the TS. This view is also supported by the prior configuration of the water molecules in the aqueous solution solvating the reactant complex and the widespread occurrence of proton relays in other processes,[55-57,59,60,64-67] including water oxidation[54,58] and enzymatic reactions.[61-63] In any event, the $\Delta G^{\ddagger}_{HT}$'s reported in Table 1 show that the reaction is viable even without involvement of any proton relay chain.

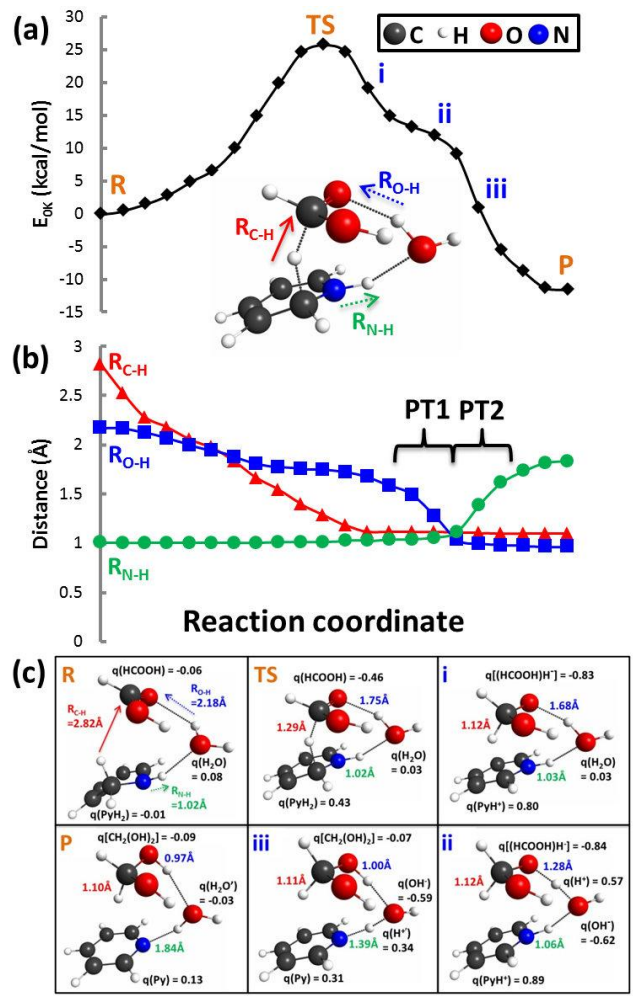

**Figure 4.** Analysis of the coupled HTPT process between **PyH₂** and HCOOH to form Py and CH₂(OH)₂ via the DHT-1H₂O model. Similar results are found for HTPT to formaldehyde. Panels: (a) energy ($E_{0K}$, not ZPE-corrected); R denotes the reactant complex, TS the transition state, i, ii, and iii are structures in the exit channel, and P, the product complex, (b) bond length, and (c) structures and charges q (calculated with the CHELPG method[141] and in the units of e) of important moieties along the reaction coordinate (corresponding to structures of R, TS, i, ii, iii and P in (a)). Both bond length and charge analyses show that the TS is dominated by HT (which is similar to the case of CO₂ reduction by **PyH₂**). Thus, the experimentally obtained -T$\Delta S^{\ddagger}_{exp}$ = 2.3 kcal/mol for a related HT reaction (eq. 1) is a good estimate for the -T$\Delta S^{\ddagger}_{HT}$ of the HCOOH reduction, despite the fact that PT is involved because PT occurs well after the HT TS, though well before the product is formed. Here, PT occurs via proton relay ~12 kcal/mol be-low (after) HCOOH's TS. This feature, as well as the absence of a TS for the PT, confirms the coupled character of the HTPT reaction. For further discussion, see the text. Because the HT and PT reactions occur in a process characterized by a single free energy TS,[142-146] we have characterized this HTPT process as *coupled*.[147] It is so distinguished from the *uncoupled* HTPT reduction of CO₂ to ultimately produce HCOOH, where first HT involving a single TS produces the HCOO⁻ intermediate, and subsequently PT to HCOO⁻ occurs independently to produce HCOOH.

To better understand how coupled HT and PT enables **PyH₂** to reduce formic acid and indeed to further support our statements above concerning its coupled character, we analyze HCOOH's reduction by **PyH₂** and its proton relay process in greater detail. In Figure 4a we show how DHT-1H₂O's energy (the internal energy $E_{0k}$ calculated at 0K and not ZPE-corrected) changes from the reactant complex (R), through the TS and structures (i, ii, and iii) energetically downhill from the TS, before ultimately reaching the product complex (P) along the computed reaction coordinate. Along the same reaction coordinate we plot the change of key bond lengths (Figure 4b). This analysis shows that the transformation from the reactant to the TS is dominated by HT. That is, $R_{C-H}$ (defined in Figure 4a) shortens from 2.82Å at R to 1.29Å at the TS while $R_{O-H}$ and $R_{N-H}$ do not change appreciably. This TS is predominantly associated with HT. Consequently, PT either to HCOOH or from oxidized **PyH₂** does not occur until well past the TS. There is no TS associated with either of these PTs, although PT does produce a shoulder in the potential energy surface ~12 kcal/mol below the TS caused by HT.

Despite the important distinction between the first two HTPT reduction steps just emphasized, the character of HCOOH's reduction by **PyH₂** is similar to that of the reduction of CO₂ in the sense that HT dominates the energetics leading to the TS for both reactions; thus, as commented upon in the caption of Figure 4, the experimental -T$\Delta S^{\ddagger}_{exp}$ value of 2.3 kcal/mol for HT from the related dihydropyridine HT reaction (eq. 1) is also a reasonable -T$\Delta S^{\ddagger}_{HT}$ estimate for HT to HCOOH by **PyH₂**.

The formic acid reduction is different from that of CO₂ in that --- as we noted above --- HCOOH's HT reaction is followed by coupled PT along the reaction coordinate, mediated by a proton relay via H-bonded water molecule(s). The first PT occurs along the exit channel ~12 kcal/mol downhill from the TS (Figures 4a and b), where the C=O oxygen of the hydroxymethanolate anion ((HCOOH)H⁻ product of HT to HCOOH) abstracts a H⁺ from its H-bonded H₂O to form methanediol and a hydroxide (OH⁻)-like moiety (characterized further below). In contrast to the CO₂ reduction case where the produced HCOO⁻ is not basic enough to initiate a proton relay, the HT intermediary product of formic acid, (HCOOH)H⁻, is sufficiently basic (pK$_a$ of methanediol is ~13)[148,149] to commence a proton relay by abstracting a proton from the neighboring H-bonded water molecule.

This first PT event (PT1) is marked by the shortening of $R_{O-H}$ from ~1.6 to ~1.0 Å. Immediately following PT1, the second PT event (PT2) occurs where the just-formed OH⁻- like moiety now abstracts a H+ from its H-bonded partner PyH+ (formed by HT from **PyH₂**) to form H₂O and more importantly, to recover the Py catalyst. This aspect of the proton relay process is marked by the lengthening of $R_{N-H}$ from ~1.0 to ~1.8 Å. This analysis clearly shows the cooperative nature of the HT and PT and that although PTs occur well into the exit channel, they act to stabilize the HT TS without participating in the configuration of the TS.

Finally, we analyze how the charges on various moieties change along the reaction coordinate. In Figure 4c it is apparent that as the reaction proceeds from R to TS the charge of **PyH₂** has become increasingly positive (q= 0.43e), while HCOOH has become increasingly negative (q= -0.46e); this is consistent with a HT reaction and correlates with the motions along the reaction coordinate shown in Figure 4b. As the hydride transfer from **PyH₂** to the HCOOH carbon becomes more complete (structure i), the (HCOOH)H⁻ moiety becomes increasingly basic (q= -0.83e) such that its carbonyl oxygen begins to abstract a proton from the H-bonded water molecule (structure ii) to form an intermediate hydroxide OH⁻ type moiety (q= -0.62e). Structure iii shows that this basic species then abstracts a proton from PyH+ and thus completes the proton relay to ultimately produce CH₂(OH)₂, while recovering the Py catalyst in the product P; H₂O' denotes a newly formed water as a result of proton relay. Figure 4 shows that **PyH₂** contains both hydridic (C₂-H) and protic (N-H) hydrogens; this is analogous to the situation for ammonia borane, which we previously showed reduces CO₂ by HTPT.[38,39]

**3.6 Third HTPT step: PyH₂ + OCH₂ → Py + CH₃OH.** We now address the third and final reduction step to produce the desired product, CH₃OH. This step follows the formation of CH₂(OH)₂, which is a hydrated formaldehyde (OCH₂). In order to effect further reduction, the sp³-hybridized CH₂(OH)₂ produced by the second HTPT must first be dehydrated to form the sp²-hybridized species OCH₂ at $K_{eq}$ ~5x10⁻⁴ (Scheme 8, route **IX**).[150] While equilibrium strongly favors the diol species, OCH₂ is significantly more reactive to HT, producing methanol via **PyH₂** + OCH₂ → Py + CH₃OH (route **X**) at low barrier, e.g. $\Delta G^{\ddagger}_{HT}$ = 6.0 kcal/mol calculated for the DHT-2H₂O model (see Table 1 for the $\Delta G^{\ddagger}_{HT}$ values and Figure 3 for the TSs). Thus this low $\Delta G^{\ddagger}_{HT}$ value suggests that the slowest step from CH₂(OH)₂ to CH₃OH is in fact likely to be the dehydration of CH₂(OH)₂ to OCH₂. The rate constant for the dehydration of CH₂(OH)₂ to OCH₂ at ambient conditions[139,140] is ~5x10⁻³ s⁻¹ (obtained at an ~neutral pH) or equivalently the estimated free energy barrier $\Delta G^{\ddagger}_{dehyd}$ is ~21 kcal/mol. Consequently, the effective rate constant for transformation of CH₂(OH)₂ to CH₃OH is that of CH₂(OH)₂ dehydration.

**Scheme 8. Dehydration of methanediol to form formaldehyde and the subsequent reduction to methanol by PyH₂.**

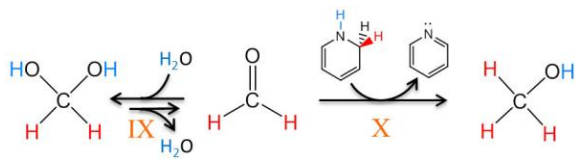

In a fashion similar to the HCOOH reduction, the reduction of OCH₂ proceeds via a *coupled* HTPT step, which we illustrate using structures determined via IRC calculations. Figure 5 shows the reactant complex R involving **PyH₂**, OCH₂ and H₂O for the DHT-1H₂O model. In this complex, the C of OCH₂ is still far from the hydridic H of **PyH₂** (e.g. $R_{C-H}$ = 2.39 Å) and all moieties are approximately charge neutral (e.g. HT has not yet commenced and all species have q ~ 0). At the TS, OCH₂ is in the process of accepting a hydride from **PyH₂** and importantly, there is no significant PT, as evidenced by the relatively large $R_{O-H}$= 1.73 Å value relative to the $R_{O-H}$ value 0.98 Å of the product. Thus, the TS consists of HT character, again justifying our use of the experimental HT activation entropy factor proposed in section 2.

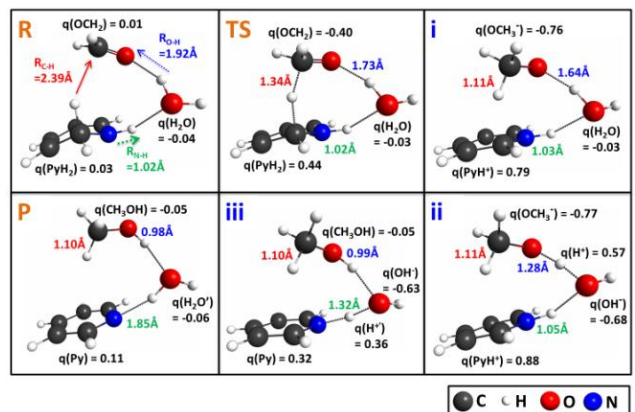

**Figure 5.** Reduction of formaldehyde by **PyH₂** to methanol (via the DHT-1H₂O model) in a coupled HTPT step. In the reactant complex R, all moieties (**PyH₂**, OCH₂, and H₂O) are approximately neutral (e.g. q ~ 0, in electronic charge units, e) and the HT reaction from **PyH₂** to OCH₂ has not commenced (e.g. $R_{C-H}$ = 2.39 Å). The reaction then proceeds to the TS, which is of HT character: OCH₂ has become more negatively charged [q(OCH₂) = -0.40e] on the way to full HT, while **PyH₂** has become more positive [q(**PyH₂**) = 0.44e], without any significant PT (e.g. $R_{O-H}$ = 1.73 Å). As the reaction progresses energetically downhill from the TS towards the product, the HT completes and methoxide anion (OCH₃⁻) is formed in structure i. The basic methoxide [q(OCH₃⁻) = -0.77e] now begins to abstract a proton from the neighboring H₂O in structure ii to form methanol (CH₃OH) in structure iii. The proton relay continues as the first PT-produced transient hydroxide anion-like OH⁻ now abstracts a proton from PyH+ to finally form the product complex P of Py, CH₃OH and H₂O', where ' denotes the water molecule newly formed as a result of the proton relay.

As the reaction progresses energetically downhill from the TS towards the product, HT completes, transiently forming the methoxide (OCH₃⁻) anion-type moiety, displayed in structure i of Figure 5. In analogy to the second HTPT step, the PT aspect of the reaction occurs well into the exit channel after the HT TS and involves no TS on the way to the reaction product. Thus, the HT and PT are *coupled* in this HTPT process. The PT aspect of the reaction involves a proton relay chain for the one and two H₂O DHT model cases. The newly formed methoxide anion-like moiety is negatively charged [q(OCH₃⁻) = -0.76e] and possesses a sufficiently basic carbonyl (pK$_a$ of methanol is ~16)[151] that it abstracts a proton from a neighboring hydrogen-bonded H$_2$O (structure ii) to initiate a proton relay cascade: a transient hydroxide anion-like moiety is produced (structure ii), which then abstracts an H$^+$ from PyH$^+$ (the oxidized **PyH$_2$** which has earlier resulted from HT) as CH$_3$OH formation is completed (structure iii), to finally form Py together with H$_2$O' and CH$_3$OH in the product complex, P. The HTPT activation free energies for the three cases are reported in Table 1. Our earlier remark about a minor activation entropy effect for the proton relay aspects of the second step also applies here.

It is worth noting that HT from a related dihydropyridine species to an aldehyde functional group has been observed experimentally.[152,153] In eq. 2, 10-methyl-9,10-dihydroacidine transfers its hydride to benzaldehyde to form benzyl alcohol in the presence of perchloric acid (HClO$_4$), which acts as the H$^+$ donor.[152] The HTPT reaction between **PyH$_2$** and OCH$_2$ to form methanol (Scheme 8, route **X**) is analogous to eq. 2; however route **X** differs slightly because **PyH$_2$** acts as both the hydride and proton donor.

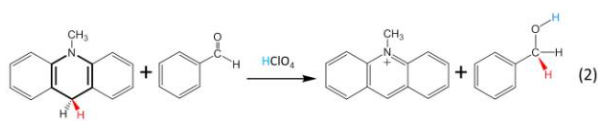

### 3.7 Commentary on the homogeneous mechanism for CO$_2$ reduction to CH$_3$OH catalyzed by pyridine.

The preceding results in this section allow us to map out a complete homogeneous mechanism of Py-catalyzed CO$_2$ reduction to CH$_3$OH via three HTPT steps (Scheme 9) where the first HTPT to CO$_2$ is uncoupled and sequential and the final two HTPT steps are coupled in character. These results are summarized in Table 1 and Figure 3. Examination of Table 1 and Figure 3 shows that the second HTPT step, that of HCOOH reduction, is the highest HTPT free energy barrier step for the reduction of CO$_2$ to CH$_3$OH by **PyH$_2$** in all cases. However, in the DHT-2H$_2$O case, the second HTPT barrier $\Delta G^{\ddagger}_{HT}$= 18.7 kcal/mol is lower than the methanediol dehydration barrier $\Delta G^{\ddagger}_{dehyd}$ of ~21 kcal/mol (see Section 3.6 and Figure 3). In this connection, it is noteworthy that substrate and/or hydride donor activation[29,136,152,154] can act to further lower $\Delta G^{\ddagger}_{HT}$. For example, K$^+$ and PyH$^+$ in solution can activate the carbonyls for HT reaction (see discussion at end of section 3.4). However, even in the absence of this additional activation, the **PyH$_2$**-catalyzed reduction of CO$_2$ to CH$_3$OH is kinetically facile. Moreover, we have found that for the second and third reduction steps, a proton relay chain can noticeably reduce the reaction barriers. However, even without these proton relays, Table 1 --- and the methanediol dehydration barrier $\Delta G^{\ddagger}_{dehyd}$ of ~21 kcal/mol --- indicate that these reactions remain viable in activation free energy terms.

**Scheme 9.** Homogeneous mechanism of Py-catalyzed CO$_2$ reduction to CH$_3$OH via PyH$_2$/Py HTPT processes.

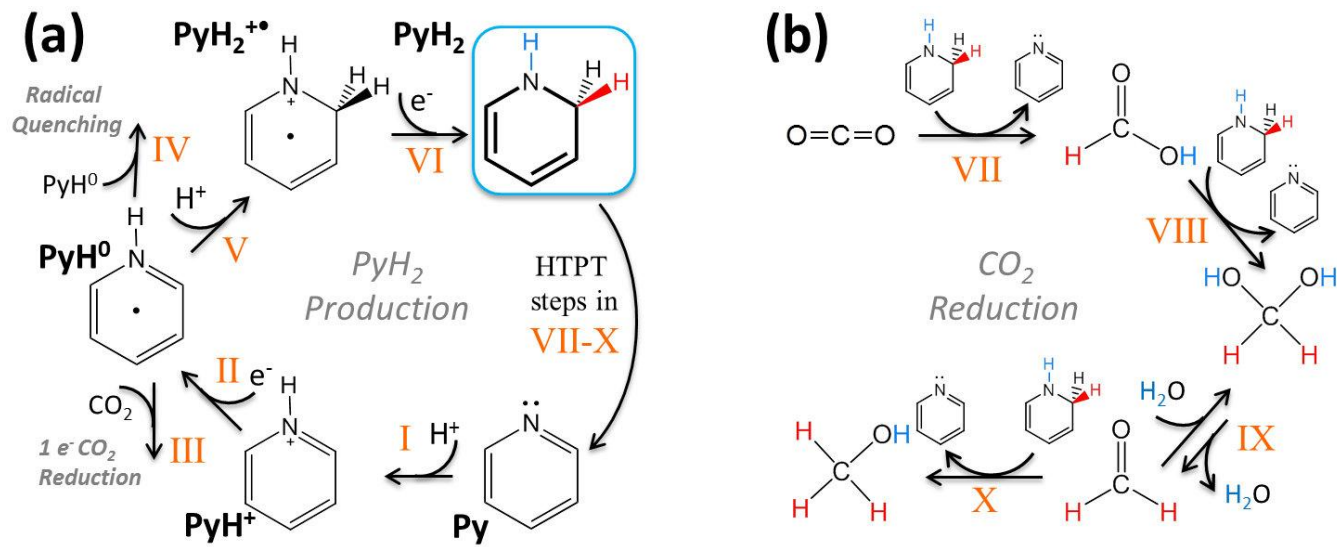

(a) **PyH$_2$** formation issues. In routes **I** and **II**,[47] Py accepts an H$^+$ to form PyH$^+$ and then an e$^-$ to form the PyH$^0$ neutral radical, which then either reduces CO$_2$ by 1 e$^-$ reduction to form PyCOOH$^0$ (route **III**)[47] or undergoes radical self-quenching (route **IV**) to produce H$_2$ + 2Py, a 4,4' coupled dimer or Py + **PyH$_2$**. Alternatively, and of most importance in the present work, in routes **V** and **VI**, PyH$^0$ accepts a second H$^+$ and then a second e$^-$ to form the potent recyclable organo-hydride **PyH$_2$**. (b) CO$_2$ reduction to methanol. In routes **VII-X**, the produced **PyH$_2$** then participates in each of three catalytic HTPT steps to reduce CO$_2$ to CH$_3$OH and H$_2$O, while recovering the Py catalyst.

For completeness, we have also considered a potential side reaction that might significantly impact the Faradaic yield for the overall **PyH$_2$**-catalyzed CO$_2$ reduction to CH$_3$OH: HT from **PyH$_2$** to a proton donor such as PyH$^+$ to evolve H$_2$ (**PyH$_2$** + PyH$^+$ = PyH$^+$ + Py + H$_2$). We have calculated that this route involves a $\Delta G^{\ddagger}_{HT}$ of 24.0 kcal/mol, which demonstrates that such unproductive heterolytic quenching to form H$_2$ is dominated by the **PyH$_2$**-catalyzed

HT to $CO_2$, HCOOH, and $OCH_2$, as well as the methanediol dehydration. The higher barrier for $H_2$ production is supported by the fact that the HT reaction by the corresponding dihydropyridine species in eq. 2a can be carried out in acidic conditions without appreciable $H_2$ production.[152] The very high (96%) Faradaic yield of the Bocarsly p-GaP system[23] is also consistent with the unfavorable heterolytic quenching to form $H_2$.

We recognize that a homogeneous pathway for pyridine-mediated $CO_2$ reduction to $CH_3OH$ has been argued to be ruled out in several recent theoretical studies,[81,91] and we briefly address this here. One key premise raised by the studies' authors is that 1e- reduction of $PyH^+$ to $PyH^0$ cannot occur at experimental conditions.[81] But this statement is not supported by the fact that highly reducing electrons are present in both the photo-electrochemical p-GaP system ($E_{CBM} \sim -1.5V$ vs. SCE at pH= 5)[74,75] and the photo-chemical $[Ru(phen)_3]^{2+}$/ascorbate system[89] to populate $PyH^+$'s LUMO ($E^0_{calc} \sim -1.3V$ vs. SCE) to form the solution phase $PyH^0$ (see the discussion in section 3.1). Another premise is that radical self-quenching will render $PyH^0$ inactive.[91] We have already pointed out in section 3.1 that radical self-quenching of $PyH^0$ can actually yield the productive $PyH_2$ via disproportionation.[104] In addition, it is relevant to note that Py-related neutral radicals of nicotinamide,[44] acridine,[45] and 3,6-diaminoacridine[46] have been experimentally observed and are key intermediate species en route to forming the related dihydropyridine species.

Finally, and in contrast to the present identification of $PyH_2$ as the important catalytic agent in homogeneous Py-mediated $CO_2$ reduction, it has been suggested that a surface-adsorbed dihydropyridine species is the active species in Py-mediated $CO_2$ reduction.[91,92] We have already noted that a solution phase dihydropyridine species is normally involved in observed HT reactions such as those in eq. 1 and 2. In any event, in our view, this surface-adsorbed species proposal does not provide a viable HT mechanism.[155] A key issue in the proposal is that the N-H bond of the adsorbed dihydropyridine is proposed to act as a hydride donor.[92] However, the N-H hydrogen is protic, not hydridic; accordingly, this suggestion is inconsistent with the extensive literature concerning HT from dihydropyridines,[29,40,41,43,76,122,123,125,126,128,152,153] including the present work, which uniformly shows that the hydride transfers from the hydridic hydrogen of the C-H bond and not from N-H.[156]

**3.8 Recovery of aromaticity drives hydride transfer from $PyH_2$.** We have shown that $CO_2$ reduction to $CH_3OH$ is accomplished via three successive HTPT steps by $PyH_2$. We now describe the principle that makes $PyH_2$ an effective HT agent. In fact, $PyH_2$'s strong hydride nucleophilicity could be regarded in a certain sense as rather surprising; it is an organo-hydride where the hydridic H is provided by a C-H bond. Consequently, $PyH_2$ differs significantly from conventional transition-metal hydrides (M-H)[22,27,127,129] in that C is more electronegative than the transition metals (M), e.g., Co, Ni and Pt. We suggest that the origin of the hydride nucleophilicity of the hydridic C-H bonds of $PyH_2$ lies in the energetics of dearomatization and aromatization of $PyH^+$,[47] a concept similar to one that has been applied to metal-ligand cooperation in catalysis involving transition-metal complexes.[157,158] During the formation of $PyH_2$, the first reduction of $PyH^+$ to $PyH^0$ dearomatizes $PyH^+$'s ring (Scheme 9a, route **II**), a destabilization consistent with $PyH^+$'s highly negative $E^0$ of $\sim -1.3V$ vs. SCE. $PyH^+$'s proclivity to regain its aromaticity drives HT from the hydridic C-H bond of $PyH_2$ to the carbon atoms of $CO_2$, HCOOH and $OCH_2$ to form reduced products and to recover the aromatic $PyH^+$ (or Py) catalyst. This mirrors the aromatization driving force several of us previously described in $PyCOOH^0$ formation via a 1e- process.[47]

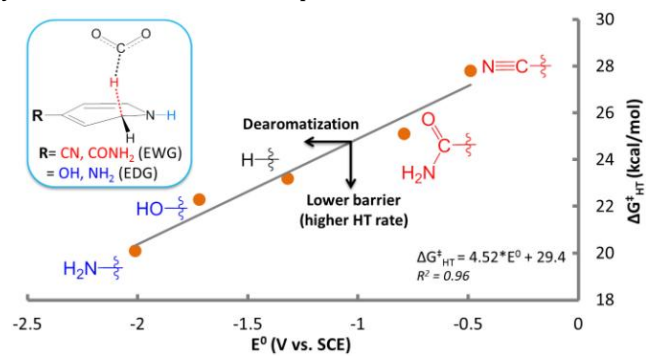

**Figure 6.** The calculated standard activation free energy 'barrier' $\Delta G^{\ddagger}_{HT}$ (kcal/mol) to hydride transfer to $CO_2$ correlates linearly with the degree of dearomatization of the hydride donor. $\Delta G^{\ddagger}_{HT}$ (kcal/mol) is calculated for hydride transfer to $CO_2$ to form $HCOO^-$ using the DHT model of Figure 2a (also shown here in the inset). $E^0$ measures the energy required to dearomatize $PyH^+$ and related protonated aromatic amines and thus serves as a quantitative measure of the degree of dearomatization. $E^0$ (V vs. SCE) is our calculated standard reduction potential for the protonated pyridine species indicated in Scheme 9a, route **II**, e.g. $PyH^+ + e^- = PyH^0$ (see SI, section 1b for details of $E^0$ calculations). We substitute $PyH_2$ with electron-withdrawing (R= CN, $CONH_2$) and electron-donating (R= OH, $NH_2$) groups in the para position of the ring to establish a wide range of $E^0$, spanning from -0.49 to -2.10V vs. SCE, and thus a broad degree of dearomatization. These groups are labeled EWG and EDG, respectively, in the Figure.

Figure 6 confirms the dearomatization-aromatization principle by demonstrating that the free energy barrier for HT to $CO_2$, $\Delta G^{\ddagger}_{HT}$, decreases with increasing cost of dearomatization as measured by the standard reduction potential $E^0$ defined in Scheme 9a, route **II**. We obtain a wide range of $E^0$ spanning from -0.49 to -2.10V vs. SCE by substituting electron-withdrawing (e.g. CN, $CONH_2$) and electron-donating (e.g. OH, $NH_2$) groups at $PyH_2$'s para position. We contend that as the $E^0$ of an aromatic species becomes increasingly negative, more energy is required to dearomatize that species by populating its LUMO (a benzene-like $\pi^*$ orbital);[159] thus $E^0$ is a quantitative measure of the energetic cost of dearomatization. The linear trend established in Figure 6 has a firm physical basis: as $E^0$ becomes more negative, the driving force to recover aromaticity increases accordingly, which in turn results in lower $\Delta G^{\ddagger}_{HT}$ values and consequently a higher hydride transfer rate. Figure 6 shows that the effect of dearomatization/aromatization on $\Delta G^\ddagger_{HT}$ enables **PyH₂** to act in its unique role as a potent hydride donor, in the present case one that catalyzes the reduction of $CO_2$ to $CH_3OH$ through three HTPT steps and which is regenerated through the **PyH₂**/Py redox couple (Scheme 9a, route **I-II-V-VI**).

### 4. Concluding Remarks

In summary, we have elucidated a kinetically and thermodynamically viable mechanism for the homogeneous reduction of $CO_2$ to $CH_3OH$ by 1,2-dihydropyridine, **PyH₂**.[160] Our proposed sequential PT-ET-PT-ET process of alternating proton and electron transfers (Scheme 9a, routes **I-II-V-VI**) that initially transforms Py into the catalytic species **PyH₂** is supported by the observation of a similar process occurring in Py-related species, e.g. nicotinamide and acridines,[44-46] where the aromatic PyH⁺ is dearomatized during the process. Subsequently, driven by the proclivity to recover aromaticity, **PyH₂** transfers three hydridic hydrogens in three successive steps to $CO_2$, HCOOH and $OCH_2$ to ultimately form $CH_3OH$ (Scheme 9b, routes **VII-X**). The initial reduction of $CO_2$ is mediated by an uncoupled, sequential HTPT process; for the subsequent HCOOH and $OCH_2$ reductions, coupled HTPT occurs, in which PT is mediated by a proton relay via one or two water molecules.

We stress that while we have theoretically demonstrated homogeneous $CO_2$ reduction proceeding after **PyH₂** formation, we do not rule out possible surface-catalyzed events, most especially on a Pt electrode. On the other hand, we suggest that both Bocarsly's p-GaP[23] and Mac-Donnell's surface-free Ru(II)/ascorbate[89] systems are homogeneous processes mediated by our proposed recyclable **PyH₂**/Py redox couple. This suggestion is reinforced by Tanaka's demonstration that the related dihydropyridine (**Ru(bpy)₂(pbnH₂)²⁺**) species homogeneously reduces $CO_2$ to $HCOO^-$ by hydride transfer;[29] in addition, the related 10-methyl-9,10-dihydroacidine has been demonstrated to convert benzaldehyde into benzyl alcohol via a HTPT step.[152] We thus theoretically predict that pyridine's intriguing catalytic behavior lies in the homogeneous HT chemistry of the **PyH₂**/Py redox couple, whose production (Scheme 9a) is driven by a dearomatization-aomatization process, as argued in connection with Figure 6.

It is noteworthy that the **PyH₂**/Py redox couple --- by its hydride transfer to carbonyl for C-H bond formation --- closely imitates the NADPH/NADP⁺ catalyzed reduction step in photosynthesis (see Scheme 5a). Our present results thus suggest that the NADPH/NADP⁺ couple is similar to the **PyH₂**/Py couple in that dearomatization is used to store energy that is subsequently used to drive HT while regaining aromaticity. Finally, we propose that the advantage of the recyclable **PyH₂**/Py redox couple extends beyond the mechanism of $CO_2$ reduction described within to provide inexpensive and green alternatives to commonly used hydride donors in organic synthesis.

## ASSOCIATED CONTENT

**Supporting Information**. Computational Methods; Overestimation of Activation Entropies using Ideal Gas Partition Functions; Thermodynamic Quantities Referenced to Reactant Complex; Recovery of the Pyridine Catalyst from the 4,4' Coupled Dimer; Linearized Poisson-Boltzmann Model of Cation Concentration Near a Biased Cathode; Reactivity of 1,2-dihydropyridine vs. 1,4-dihydropyridine Towards $CO_2$; Hydride Transfer from the N-H Bond of 1,4-dihydropyridine; Coordinates of Molecular Structures. This material is available free of charge via the Internet at http://pubs.acs.org.

## AUTHOR INFORMATION

**Corresponding Author**

\* Charles B. Musgrave (charles.musgrave@colorado.edu)

## ACKNOWLEDGMENT

This work was supported in part by NSF Grant CHE-1112564 (JTH). We appreciate support from the University of Colorado Boulder and the Center for Revolutionary Solar Photoconversion of the Colorado Renewable Energy Collaboratory (CBM). CBM and JTH are a Fellow and Affiliate, respectively of RASEI and CBM is a Fellow of the Materials Science Program of the University of Colorado Boulder. We also gratefully acknowledge use of the supercomputing resources of the Extreme Science and Engineering Discovery Environment (XSEDE), which is supported by National Science Foundation (NSF) grant number ACI-1053575. This work also utilized the Janus supercomputer, which is supported by NSF grant number CNS-0821794 and the University of Colorado Boulder. We also thank Dr. Yong Yan of NREL for useful discussions.